\documentclass[conference]{IEEEtran}
\IEEEoverridecommandlockouts

\usepackage{cite}
\usepackage{amsmath,amssymb,amsfonts}
\usepackage{algorithmic}
\usepackage{graphicx}
\usepackage{textcomp}
\usepackage{xcolor}
\usepackage{url}
\usepackage{dblfloatfix}   

\def\BibTeX{{\rm B\kern-.05em{\sc i\kern-.025em b}\kern-.08em
    T\kern-.1667em\lower.7ex\hbox{E}\kern-.125emX}}
\begin{document}

\title{Full State-Space Visualisation of the 8-Puzzle: Feasibility, Design, and Educational Use}

\author{
\IEEEauthorblockN{Ian Frank}
\IEEEauthorblockA{
\textit{Dept. Complex and Intelligent Systems} \\
\textit{Future University Hakodate} \\
Hakodate, Japan \\
ianf@fun.ac.jp
}
\and
\IEEEauthorblockN{Kanata Kawanishi}
\IEEEauthorblockA{
\textit{Dept. Media Architecture} \\
\textit{Future University Hakodate} \\
Hakodate, Japan \\
kawakana0510@gmail.com
}
}

\maketitle

\begin{abstract}
  Search algorithms are a foundational topic in artificial
  intelligence education, yet even simple domains can generate large
  state spaces that challenge learners’ ability to form accurate
  mental models. This paper presents an interactive learning system
  that demonstrates the feasibility of visualising the \emph{entire}
  reachable state space of the 8-puzzle (181,440 states), while
  tightly coupling abstract graph structure with concrete puzzle
  manipulation. Built using Unity and modern GPU-based rendering
  techniques, the system enables real-time exploration of global
  structure, step-by-step execution of search algorithms, and direct
  comparison of how different strategies traverse the same space. We
  describe the system’s design, visualisation layouts, and educational
  use, reporting findings from an initial classroom deployment and
  pilot study with students at different levels of university
  education. Overall, the results indicate that full state-space
  visualisation is both technically feasible and educationally
  valuable for supporting conceptual understanding of search behaviour
  within this canonical problem domain.
\end{abstract}


\section{Introduction}

Search algorithms such as breadth-first search (BFS), depth-first search
(DFS), and heuristic search are foundational topics in artificial
intelligence education. Despite their centrality, learners frequently
struggle to form accurate mental models of how these algorithms operate
within large state spaces. Instruction commonly introduces search using
tree or graph representations, but these are typically limited to small,
abstract examples. As problem size grows beyond toy cases, the search
space quickly becomes too large to visualise directly, leaving learners
to imagine structure and scale for themselves.

The 8-puzzle has served as a canonical benchmark problem in artificial
intelligence since at least the early work of Doran and Michie on
graph traversal programs in the 1960s~\cite{DoranMichie1966}. Its
appeal lies in the combination of simple, easily understood rules with
a non-trivial underlying search space. Although the puzzle is simply
described informally, its reachable state space comprises 181,440
distinct configurations connected by legal moves, forming a highly
structured graph.

Historically, this scale placed full visualisation of the state space
beyond the reach of both research and educational tools.
As a result, the 8-puzzle has typically been used to illustrate search
algorithms through partial trees, execution traces, or isolated
solution paths, as seen both in standard AI textbooks and in interactive
web-based solvers~\cite{RussellNorvig2021,EightPuzzleSolverWeb}.
Recent work~\cite{FrankKawanishiJSET} suggested that advances in
commodity hardware and real-time graphics now make it feasible to
generate and interactively visualise the complete state space using
accessible tools. Building on this demonstration of technical
feasibility, the present paper focuses on the design, interaction, and
educational use of a system that exposes the entire search space to
learners.

The interactive learning system that we present visualises the entire
reachable state space of the 8-puzzle and tightly couples abstract
graph structure with concrete puzzle manipulation. By allowing
learners to move seamlessly between global state-space structure and
individual puzzle transitions, the system supports exploration of
search depth, connectivity, and differences in algorithm behaviour for
standard search methods such as BFS, DFS, and A*. In doing so, the
work produces a novel coupling between concrete puzzle interaction and
abstract full search-space representation.  To support adoption and
reuse in educational settings, the system is publicly available
online, accompanied by background material and demonstration content
intended for instructors and students.

The remainder of this paper is organised as follows. Section~II
reviews related work in algorithm visualisation, search education, and
puzzle-based state-space representations, and identifies gaps
addressed by this study. Section~III describes the system design,
focusing on the technical challenges of rendering and interacting with
a complete state space at scale. Section~IV details the visualisation
and interaction design, including multiple layout strategies and the
coupling between concrete puzzle manipulation and abstract structure.
Section~V presents educational use cases illustrating how the system
is employed in instructional settings, followed by an initial user
evaluation in Section~VI. The paper concludes with a discussion of
implications and directions for future work.

\section{Related Work}

Algorithm visualisation has a long history in computer science education,
with early work demonstrating the potential to support understanding of
dynamic processes that are difficult to grasp from static text alone.
Surveys of the field, however, have repeatedly noted a strong imbalance
in focus: the majority of visualisation systems address
sorting algorithms and simple data structures, with comparatively few
tools dedicated to search algorithms or state-space exploration
\cite{Shaffer2010,Naps2003}.

Existing visualisations of search algorithms typically emphasise
procedural execution. Many systems illustrate breadth-first search,
depth-first search, or A* by animating node expansion, frontier
growth, and solution paths in trees or graphs. Such tools are
effective for explaining algorithm mechanics, and are widely used in
introductory computer science courses, including maze-based
pathfinding demonstrations and interactive tutorials such as those by
Red Blob Games~\cite{RedBlobGames}. However, these representations are
necessarily local: they show only a small portion of the search space
at any given time, and provide limited support for understanding
global structure, scale, or connectivity.

A smaller body of work has explored whole-space representations of
search domains. For the 8-puzzle in particular, radial or
sunburst-style visualisations have been used to display complete
search trees or to analyse heuristic behaviour~\cite{Hatem2009}. These
studies demonstrate the potential of large-scale state-space
visualisation, especially for examining properties such as solution
depth and heuristic error. However, such visualisations are typically
static or precomputed, and do not support interactive navigation or
direct manipulation of puzzle states.

Research on interactivity in algorithm visualisation further complicates
the picture. While increased interactivity can encourage exploration
and hypothesis testing, several studies have shown that more
interactivity does not necessarily lead to better learning outcomes
\cite{Naps2003}. In particular, systems that focus heavily on local
manipulation risk fragmenting learners’ understanding, making it
difficult to maintain a coherent sense of global structure. These
findings highlight the importance of balancing exploratory freedom with
stable, interpretable representations.

The 8-puzzle has also played a central role in research on heuristic
search. Foundational work on admissible heuristics, including
Manhattan distance and pattern databases, has focused on algorithmic
efficiency and optimality rather than on learner-facing
representations \cite{Culberson1998,Korf2000}.  Similarly, graph
drawing and layout techniques based on forces have been extensively
studied in information visualisation, but examples of their
application to complete state spaces in interactive educational tools
remain limited.

Taken together, prior work reveals several persistent gaps. Few systems
support interactive exploration of an entire search space at meaningful
scale; fewer still integrate global state-space representations with
concrete problem manipulation; and existing educational tools tend to
prioritise procedural execution over structural understanding. The
system presented in this paper addresses these gaps by combining
full-state-space visualisation, multiple layout strategies, and direct
coupling between puzzle operations and abstract search-space structure
within a single interactive environment.

\section{System Design}

The gaps identified in prior work point to a combined technical and
representational challenge. Supporting interactive exploration of an
entire search space requires not only scalable rendering and layout
computation, but also visual structures that allow learners to perceive
global organisation while retaining access to local transitions. The
system described here was designed to address these challenges by making
full-state-space visualisation both technically feasible and
educationally usable.

The system is implemented using Unity and targets both desktop execution
and deployment via WebGPU-enabled web browsers. The complete reachable
state space of the 8-puzzle was precomputed in advance, resulting in
181,440 nodes and 241,920 edges. Each node represents a unique tile
configuration, and edges correspond to legal puzzle moves. This static
graph structure serves as the basis for multiple interactive layout and
visualisation modes.

\begin{figure}[t]
  \centering 
  \includegraphics[width=\linewidth]{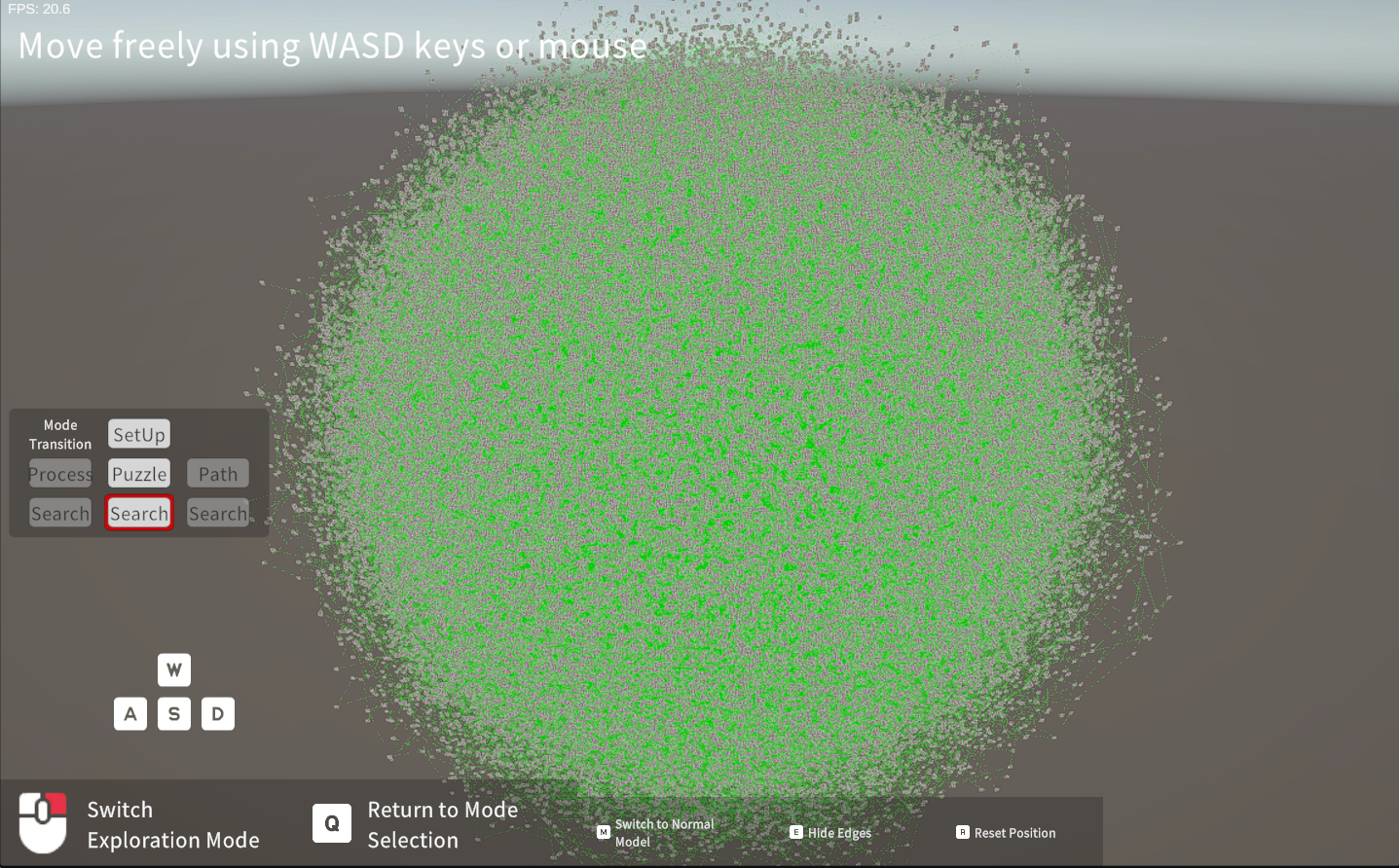}
  \caption{Force-based, zoomable layout of the complete 8-puzzle state space
                (181,440 states), rendered interactively in real time using GPU instancing,
                illustrating the global structure of the space as shaped by state connectivity.}
  \label{fig:force}
\end{figure}

Rendering and interacting with a graph of this scale in real time
poses significant technical challenges. Na\"ive approaches to graph
rendering quickly exceed practical limits, leading to poor frame rates
and unresponsive interaction. To address these constraints, the system
employs GPU instancing and custom shaders, allowing large numbers of
nodes with shared visual properties to be rendered efficiently in a
batched manner. This enables smooth zooming, panning, and selection
even when the full state space is visible (Figure~\ref{fig:force}).

Layout computation presents a second major challenge. Among the
layouts supported by the system, the force-based layouts play a
central role in conveying global structure. In this mode, nodes are
positioned on a spherical surface using simulated attractive and
repulsive forces, producing a spatial arrangement that supports a
global overview of connectivity while reducing edge crossings and
visual clutter. Computing and maintaining such layouts for hundreds of
thousands of nodes motivates the use of efficient parallel
computation. The system therefore employs Unity’s Job System and Burst
Compiler to structure layout calculations, supporting responsive
interaction even at this scale on standard consumer hardware.

Because the system is intended for educational use across a range of
devices and teaching contexts, several mechanisms are provided to manage
visual complexity. These include layout options that enforce minimum
spacing between nodes to reduce overlap, the ability to toggle edge
visibility, and the use of lightweight node models by default, with
optional switching to more detailed representations. These choices give
instructors and learners control over visual density and rendering
costs, establishing a practical baseline for full-scale, interactive
state-space visualisation without overwhelming either the hardware or
the user.

\section{Visualisation and Interaction Design}

The visualisation and interaction design of the system is guided by the
need to support both global and local understanding of search behaviour.
Rather than relying on a single representation, the system provides
multiple complementary layout strategies, each highlighting different
structural properties of the search space. Learners can switch between
layouts dynamically, allowing the same underlying state space to be
examined from multiple conceptual perspectives.

\subsection{Force-Based Layout}

The force-based layout is designed to support a global overview of the
search space, with a particular emphasis on conveying overall scale and
extent. Nodes are positioned using a lightweight force-based placement
heuristic, combining repulsive forces to promote dispersion with
attractive forces derived from state transitions to prevent uniform
random scattering. The resulting spatial arrangement does not aim to
faithfully reproduce exact structural or metric relationships within
the state space, but instead provides a visually coherent distribution
that allows learners to perceive the size and spread of the domain
beyond individual solution paths.

\subsection{Depth-Based Layout}

In the depth-based layout, nodes are arranged concentrically according to
their search depth from a selected initial state. This representation makes
depth relationships explicit and provides a clear visual account of how the
search frontier expands over time. Because nodes at the same depth are
grouped spatially, learners can readily perceive the breadth and
redundancy of exploration at different stages of the search.

This layout is particularly effective for comparing uninformed search
strategies. When BFS and DFS are visualised in this representation,
differences in exploration order, frontier growth, and revisiting of
states become immediately apparent.  As a result, learners can relate
algorithmic control flow to spatial structure in the search space,
rather than relying solely on code-level descriptions or execution
traces.

\subsection{Heuristic-Distance Layout}

\begin{figure}[t]
  \centering
  \includegraphics[width=\linewidth]{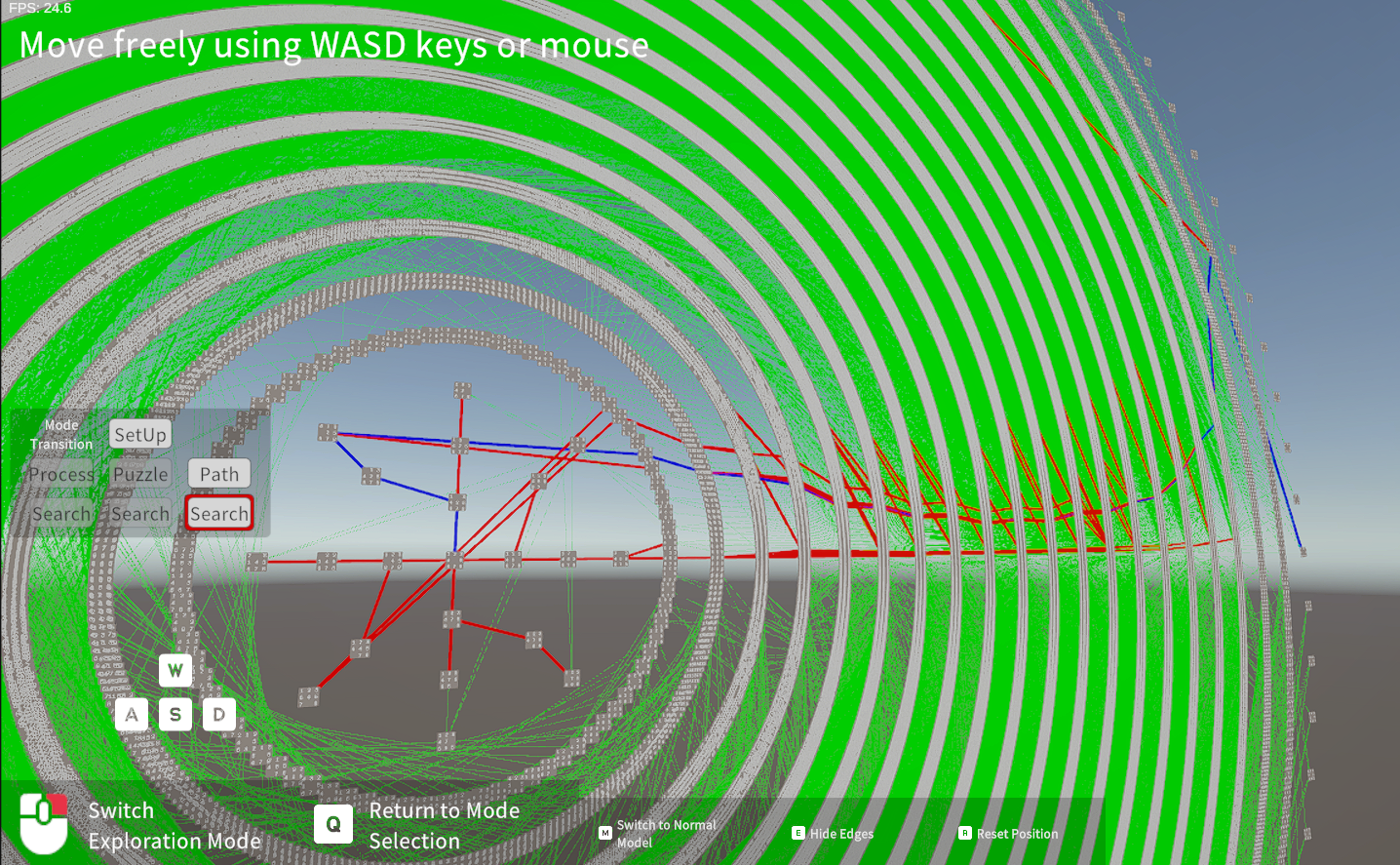}
  \caption{Heuristic-distance layout of the complete 8-puzzle state
    space, illustrating A* search. Nodes are arranged concentrically
    according to their Manhattan distance from the goal state. The
    highlighted blue path shows the sequence of states explored by A*
    from a given start configuration to the goal, illustrating how
    heuristic information reshapes traversal behaviour.}
  \label{fig:heuristic}
\end{figure}

The heuristic-distance layout positions nodes according to their
Manhattan distance from the goal state, forming concentric layers that
reflect estimated proximity to a solution. This representation makes
heuristic structure explicit and supports reasoning about informed
search behaviour. When A* search is visualised in this layout, learners
can observe how heuristic guidance shapes exploration of the state
space, in contrast to uninformed strategies such as breadth-first and
depth-first search operating over the same graph (Figure~\ref{fig:heuristic}).

\subsection{Interaction and Concrete--Abstract Coupling}

Interaction is central to the system’s educational design. Learners can
navigate the three-dimensional state space, moving between global
overviews and local neighbourhoods to examine structural relationships.
In a separate interaction mode, the puzzle interface allows direct
manipulation of tile configurations, with each move synchronised to the
corresponding transition in the state-space visualisation. This design
enables learners to alternate between abstract exploration and concrete
puzzle interaction while maintaining a clear correspondence between
local actions and global structure.

Search algorithms can be executed step by step or automatically, with
traversal paths highlighted across the entire state space. By observing
how different algorithms traverse the same graph under different
layouts, learners are encouraged to compare search strategies in terms
of structure and behaviour rather than solely by performance metrics. This
tight coupling between representation, interaction, and algorithm
execution distinguishes the system from prior visualisation tools and
supports exploratory, hypothesis-driven learning.

Together, these visualisation and interaction techniques are intended
to support open-ended, comparative engagement with search algorithms
rather than prescriptive instruction. By allowing learners to move
between representations, manipulate the puzzle directly, and observe
algorithm behaviour across the same underlying state space, the system
encourages hypothesis formation and reflection. The following section
describes how these capabilities are used in educational settings and
illustrates typical learning activities supported by the system.

\section{Educational Use Cases}

The system has been designed for use in undergraduate artificial
intelligence courses covering uninformed and informed search. It is
publicly available as a browser-based application at
\url{https://8puzzle.ianlab.org}, requiring no local installation or
specialised software. Rather than replacing existing instructional
materials, the system is intended to complement lectures, textbooks,
and programming exercises by making the structure of search spaces
directly observable. The system’s browser-based deployment allows
students to revisit concepts at their own pace, reinforcing
understanding through repeated, self-directed interaction.

This deployment model aligns with long-standing recommendations in the
algorithm and program visualisation literature, which emphasise
platform independence, ease of access, and low barriers to adoption as
critical factors for effective educational use~\cite{Naps2003}. By
eliminating setup overhead and enabling immediate use in both classroom
and self-directed contexts, the system supports flexible integration
into existing courses.

One common use case is comparative exploration of search strategies.
After encountering BFS, DFS, and A* in lectures, students can be
invited to execute the algorithms on the same 8-puzzle instance within
the system, revealing how algorithms differ in terms of
characteristics such as search depth, redundancy, and
goal-directedness.

A second use case focuses on developing intuition for
heuristics. Using the heuristic-distance layout, students can examine
how Manhattan distance imposes structure on the state space and influences A*
exploration.  Rather than treating heuristics as abstract cost
functions, learners can see how heuristic values correspond to regions
of the graph and how these regions shape the search trajectory. This
visual perspective supports discussion of admissibility, efficiency,
and the limits of heuristic guidance.

The system also supports exploratory, hypothesis-driven activities.
Students can manipulate puzzle states directly, predict how a given
algorithm will traverse the space, and then test these predictions by
running the algorithm and observing the resulting paths. Because the
full state space remains accessible, learners are encouraged to
reflect on both local decisions and their global consequences.

\section{Formative Evaluation of Educational Use}

We obtained initial feedback on the system’s use in practice via
a formative evaluation combining a small pilot study with
informal classroom use. The aim was not to measure learning outcomes
quantitatively, but to assess whether whole-state-space visualisation is
understandable, engaging, and perceived as useful by students with
varying levels of prior knowledge.

\subsection{Participants and Procedure}

The pilot study was conducted with ten students from our home
university, ranging from first-year undergraduates to first-year
master’s students.  Seven participants had not yet formally studied
search algorithms. After a brief introduction, participants interacted
with the system and completed a questionnaire including Likert-scale
items and free-text responses.

In addition, the system was briefly introduced before the winter break
of an undergraduate AI class, where a larger group of students
interacted with it informally. A total of 102 students provided short
written comments and a simple usefulness rating based on a short
hands-on experience.

\subsection{Quantitative Findings Overview}

In the pilot study, responses to the question “Did you enjoy learning
with this system?” showed no negative evaluations: no participant
selected “disagree” or “strongly disagree.” Responses were distributed
across “strongly agree,” “agree,” and “neutral,” suggesting that the
system was broadly accepted even by students without prior exposure to
search algorithms.

A System Usability Scale (SUS) questionnaire \cite{Brooke1996SUS}
yielded an average score of 58.75. While this value is below commonly
cited benchmarks, it should be interpreted cautiously. Many
participants had not yet studied search algorithms and so interacted
with the system without being fully equipped to understand its
purpose. As such, this result primarily establishes a baseline for
future improvement of instructional framing and onboarding support,
rather than indicating fundamental limitations. This
interpretation is consistent with the positive qualitative findings
reported in the following subsection.

\begin{figure}[bt]
  \centering 
  \includegraphics[width=\linewidth]{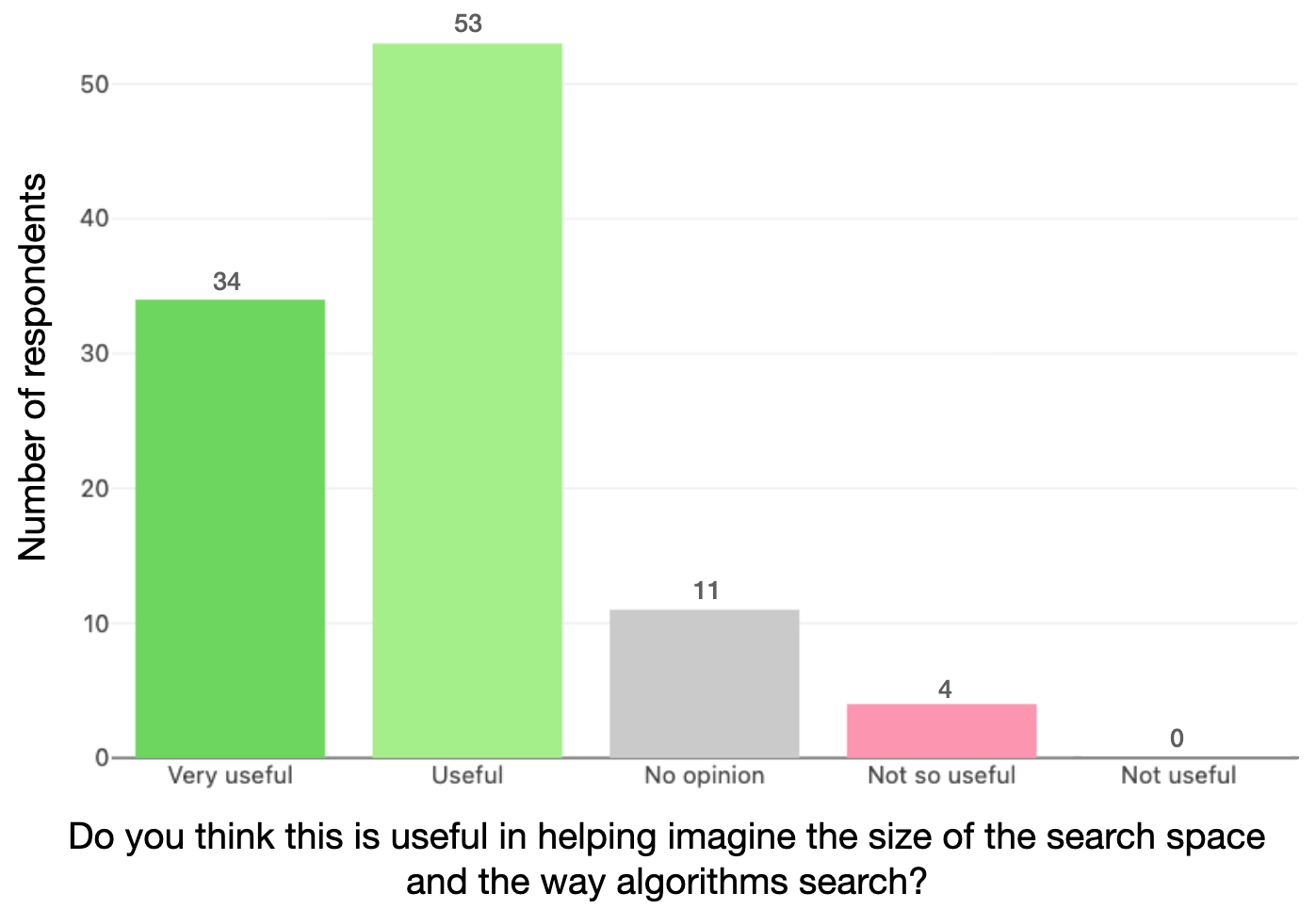}
  \caption{Student responses on a five-point Likert scale following a  
    brief in-class use of the system, showing perceived usefulness for  
    intuitively understanding the size of the search space and the  
    behaviour of search algorithms. Responses were collected after a  
    short hands-on interaction at the end of an undergraduate AI  
    lecture (N = 102).}
  \label{fig:eval}
\end{figure}

The perceived usefulness data from the larger-scale, in-class
deployment is shown in Figure~\ref{fig:eval}, which represents the
responses to a brief usefulness question administered after a short
hands-on interaction at the end of an undergraduate AI
lecture. Possible answers were on a five-point Likert scale ranging
from “not useful” to “very useful”; no participants selected the
lowest category. The results indicate that the majority of students
judged the system to be useful or very useful for understanding search
and search spaces.

\subsection{Qualitative Findings}

Qualitative feedback collected across both the pilot study and the
classroom intervention revealed several consistent themes. Many
students reported gaining an intuitive understanding of the scale of
the search space, often expressing surprise at its size despite the
simplicity of the puzzle rules. One participant commented directly that they
“didn’t expect that just eight numbered tiles and one blank could
produce so many possible states,” while others also noted that the
visualisation was key for grasping the scale and complexity.

Students also highlighted the value of spatial navigation and movement
for understanding abstract search processes. Examples include one 
report that ``the interactive view makes the search process much
easier to imagine than code or static figures,'' and another
response that ``the way breadth-first search spread
out in a circular pattern was visually pleasing''.

Comparative understanding of search strategies emerged as another
recurring theme. Learners reported that visualising multiple
algorithms within the same state space helped clarify qualitative
differences in behaviour, as reflected in comments such as ``It showed
that the way of moving toward the solution differs depending on the
algorithm'', and ``This visualisation helps intuitively understand
how... algorithms expand states step by step. It is especially helpful
for comparing different search strategies and their efficiency.''

At the same time, participants identified areas for improvement.
Comments noted that dense edge structures could make individual paths
hard to follow, and that some colour encodings were difficult to
distinguish.  One student innovatively suggested that adding sound
effects might increase the sense of immersion.  These
observations inform ongoing refinement of visual density, path
emphasis, and visual semantics.

\section{Discussion and Implications}

The formative evaluation suggests that whole-state-space visualisation
can support aspects of conceptual understanding that are difficult to
achieve through traditional instructional materials alone. In
particular, student feedback repeatedly emphasised improved intuition
for the scale and structure of the search space—an aspect that is often
left implicit once problems exceed toy examples.

Comments highlighting the value of spatial movement and synchronised
puzzle interaction support the central design motivation of this work:
that search behaviour is not only procedural but spatial, and that
linking concrete actions with abstract structure can complement
step-by-step explanations. Seeing multiple algorithms within a shared
state space also creates opportunities for qualitative comparison of
traversal behaviour, redundancy, and goal-directedness, encouraging
forms of understanding that go beyond purely performance-driven
evaluation.

At the same time, the feedback points to important limitations. Visual
density, colour semantics, and path distinguishability strongly affect
interpretability in large graphs, and the modest SUS score observed in
the pilot study highlights the need for clearer instructional framing
and onboarding, particularly for learners new to search algorithms.

From a practical perspective, the system’s browser-based,
installation-free deployment supports low-friction adoption in
educational settings and follows established recommendations for
educational visualisation tools. The work also aligns with
discussions of digital transformation (DX) in education that
emphasise learning through changes in workflows and practice, in
which understanding develops through sustained interaction with
systems and artefacts~\cite{FrankPCF11}. Viewed in this way, the system
functions not only as a visualisation aid, but as a medium through
which learners and educators can engage directly with the structure
and behaviour of search algorithms.

\section{Conclusion and Future Work}

This paper has presented an interactive learning system that visualises
the entire state space of the 8-puzzle and tightly links abstract search
structure to concrete puzzle manipulation. By making a large and highly
connected search space navigable and explorable, the system supports a
more intuitive understanding of core AI concepts, including search
depth, heuristic guidance, and algorithm behaviour.

Future work will focus on strengthening the system’s role as an
educational resource through improved instructional framing, clearer
support for algorithm comparison, and expanded formative evaluation.
Additional effort will be directed toward developing guidance and
instructional materials to help educators integrate the system
effectively into AI courses. Rather than broadening the scope to
additional problem domains, the emphasis will be on maintaining and
evolving the existing system to support sustained use in educational
settings.

\bibliographystyle{IEEEtran}
\bibliography{frank}

\end{document}